\documentclass[nofootinbib,
article,pre,
twocolumn,
eqsecnum,
amsmath,amssymb,
amsmath,amssymb, aps,
]{revtex4}
\usepackage[switch]{lineno}

\usepackage{lineno} 
\usepackage[dvipsnames]{xcolor} 
\usepackage{epsfig}
\usepackage{amsfonts}
\usepackage{amsmath}
\usepackage{wasysym}
\usepackage{amssymb}
\usepackage{bm} 
\usepackage[framemethod=PStricks]{mdframed}
\usepackage{lipsum}
\usepackage{float}
\usepackage{soul}
\usepackage[normalem]{ulem}

\usepackage{pdflscape}
\usepackage{graphicx}
\usepackage{ulem}
\newcommand{\be}{\begin{equation}}
\newcommand{\ee}{\end{equation}} 
\newcommand{\bea}{\begin{eqnarray}} 
\newcommand{\eea}{\end{eqnarray}}

\usepackage[utf8]{inputenc}
\usepackage{bbm} 				% identity operator
\usepackage{epstopdf}				% self explanatory
\usepackage{verbatim}    			% long comments
\usepackage{sidecap}                % captions on the side
\usepackage{indentfirst}

\definecolor{lightred}{rgb}{1.0, 0.13, 0.32}

\newcommand{\mean}[1]{\left\langle#1\right\rangle}
\newcommand{\lr}[1]{\left(#1\right)}
\newcommand{\tg}{\tilde\Gamma}

\usepackage[utf8]{inputenc}
\usepackage[T1]{fontenc}
\usepackage{color}
\usepackage{float}
\usepackage{soul}
\usepackage[dvipsnames]{xcolor}
\usepackage{braket}
\newcommand\bes{\begin{equation*}}
\newcommand\ees{\end{equation*}}

\newcommand\refp[1]{({\ref{#1}})}

\definecolor{armygreen}{rgb}{0.0, 0.5, 0.0}

\newcommand{\vbib}[5]{{#1}, {#2} \textbf{{#3}}, {#4} ({#5}).}

\usepackage{amssymb}
\usepackage{amsmath}
\usepackage{amsthm}
\usepackage{hyperref}

\begin{document}

\title{Vortex polarization and circulation statistics in isotropic turbulence}

\author{L. Moriconi$^{1}$, R.M. Pereira$^{2}$, and V.J. Valad\~{a}o$^{3}${\footnote{Corresponding author: victor.dejesusvaladao@unito.it}}}
\affiliation{$^{1}$Instituto de F\'{i}sica, Universidade Federal do Rio de Janeiro, C.P. 68528, CEP: 21945-970, Rio de Janeiro, RJ, Brazil}
\affiliation{$^{2}$Instituto de F\'{i}sica, Universidade Federal Fluminense, 24210-346 Niter\'{o}i, RJ, Brazil}
\affiliation{$^{3}$Dipartimento di Fisica and INFN - Università degli Studi di Torino, Via Pietro Giuria, 1, 10125 Torino TO, Italy.}
%Dipartimento di Fisica and INFN, Università degli Studi di Torino, via P. Giuria 1, 10125 Torino, Italy}
%\date{May 2020}
\vspace{1.5cm}

\begin{abstract}
\noindent We carry out an in-depth analysis of a recently introduced vortex gas model of homogeneous and isotropic turbulence. Direct numerical simulations are used to provide a concrete physical interpretation of one of the model's constituent fields: the degree of vortex polarization. Our investigations shed light on the complexity underlying vortex interactions and reveal, furthermore, that despite some striking similarities, classical and quantum turbulence exhibit distinct structural characteristics, even at inertial range scales. Crucially, these differences arise due to correlations between the polarization and circulation intensity within vortex clusters. 

\end{abstract}

\maketitle

%\date{}

\section{Introduction}\label{sec1}

Turbulence, a fascinating and intricate phenomenon, is a natural state of fluid motion that pervades our world. 
A deeper understanding of turbulence is expected to lead to technological advancements in fields such as aircraft turbulence prediction \cite{cat}, turbulent combustion \cite{combust}, and automotive design \cite{auto}. A milestone in this endeavor is due to Kolmogorov in 1941 (K41) \cite{k41}, who, with a self-similar approach to the velocity structure function, made predictions which found experimental evidence decades later \cite{obsk41}. However, it is now widely recognized that the K41 theory does not offer a definitive representation of turbulence. This limitation arises from significant non-Gaussian deviations of turbulent fluctuations \cite{frisch}, often associated with the tendency exhibited by turbulence to self-organize into thin elongated vortex structures carrying an expressive amount of kinetic energy \cite{kaneda}.

In spite of its central importance in fluid dynamics, velocity circulation was historically overlooked in the context of turbulence research. It was more than 50 years after the publication of the K41 theory that the circulation variable was first explored by Migdal \cite{migdal1} in the early 1990s. In the past decade, a remarkable improvement in hardware and software platforms enabled the exploration of high Reynolds number direct numerical simulations (DNS) of the Navier-Stokes equations. This has naturally renewed interest in circulation and the analysis of its statistical properties \cite{iyer1,iyer2}, which in turn encouraged progress both from the theoretical \cite{migdal2} and the modeling \cite{apol1,mori1,mori2,mori3,apol2,mori4} points of view. The apparent simplicity of this variable also pushed forward the computational and experimental analyses of circulation in other systems such as quantum \cite{muller1,muller2} and quasi-two-dimensional turbulence \cite{zhu,muller3}.

A promising phenomenological model of circulation statistics, referred to as vortex gas model (VGM) \cite{apol1}, combines a mathematically formal model of the cascade nature of turbulence through the Gaussian multiplicative chaos (GMC) \cite{vargas} with the structural view of turbulence, seen as an entangled gas of vortex filaments. The resulting statistical model not only accurately reproduces the general statistical behavior of velocity circulation observed in DNS, but motivated the discovery of new phenomena, as, for instance, a statistical repulsion between vortex fluid structures at small scales \cite{mori3}. In the context of probability distribution functions (PDFs), VGM gives a fundamental interpretation of the behavior of extreme far tails through the breakdown of multifractality \cite{mori2,mori4}, and offers a straightforward explanation of the multifractal behavior of circulation in terms of a maximum vortex packing fraction inside a fixed contour $\mathcal{C}$ where the circulation is computed \cite{thesisvic}.

The VGM is described in terms of two statistically independent fields. One field entails the GMC framework needed to model intermittent energy fluctuations, akin to the Obukhov-Kolmogorov modeling (OK62) \cite{O62,K62}, and its physical role has been scrutinized in \cite{mori4}. The other field is self-similar, targeting the K41 scaling exponents. The present study aims to validate, from statistical analyses of DNS data, this self-similar modeling field of the VGM. By adapting the approach of cluster summation in quantum turbulence \cite{muller2} and investigating the spatial correlation function of detected structures on DNS data, we unveil clear distinctions between classical and quantum turbulence. Additionally, our findings align with the GMC model and an analogous $4/5$-law for circulation statistics.

This paper is organized as follows. Sec.~\ref{sec2} outlines the main ideas of the VGM and its consequences to the cluster summation which are additionally strengthened in Sec.~\ref{sec3} through the analysis of high Reynolds number DNS data. In Sec.~\ref{sec4} we summarize our findings, in contrast to the current understanding of cluster summation in the context of quantum turbulence.

\section{Vortex Gas Model and Cluster summation}\label{sec2}

The VGM was first introduced in \cite{apol1} and further developed in \cite{mori1,mori3,mori4}, where circulation is accounted for at planar domains by the use of Green's theorem, as in
\be
\Gamma_r = \int_{\mathcal{D}_r}\!\bm{\omega}\,d^2x \, ,
\ee
where $\mathcal{D}_r$ is the surface delimited by some contour of typical linear size $r$ and $\bm{\omega}$ is the vorticity.
A sum of dilute, quasi-homogeneously distributed point vortices composes the statistical framework of the model. Moreover, at inertial range scales, the statistical properties of $\Gamma_r$ are encoded using two basic constituents, as previously anticipated,
\be\label{circ}
\Gamma_r=\xi_r\int_{\mathcal{D}_r}d^2x \ \tilde\Gamma(\bm{x})\ .\
\ee
The variable $\xi_r$ is the coarse-grained squared-root dissipation rate,
\be\label{qi}
\xi_r=\frac{1}{\eta_K^2}\int_{\mathcal{D}_r}d^2x \sqrt{\frac{\varepsilon(\bm{x})}{\mean{\varepsilon}}}\ ,\
\ee
modeled as a bounded GMC field, while $\varepsilon(\bm{x})$ is the local energy dissipation rate and $\eta_K$ the usual Kolmogorov scale
\cite{mori1,mori4}. In addition, $\tilde\Gamma(\bm{x})$ is a Gaussian random field with zero mean and pair correlation function
\be\label{1dot25}
\mean{\tg(\bm{x}) \tg(\bm{y})}\sim|\bm{x}-\bm{y}|^{-\alpha} 
\ee
at inertial range scales (properly regularized for small scales $|\bm{x}-\bm{y}|\ll\eta_K$). Using Eqs.~(\ref{circ}-\ref{1dot25}), it is not hard to show that $\mean{|\Gamma_r|^p}\sim r^{\lambda_p}$ with
\be\label{lambdap}
\lambda_p=\frac{(4-\alpha)}{2}p+\frac{\mu}{8}p(1-p)\ ,\ 
\ee
where $\mu\approx0.17$ is the intermittency parameter related to the spatial decay of the dissipation correlation function \cite{tang}. The above equation is only valid for low-order exponents $p\lesssim6$. For higher-order statistics, the linearization effect is accounted for by the bound imposed on the GMC field \cite{mori1,mori4}.

A deeper understanding of the Eq.~\refp{qi} was provided by \cite{mori4}, where the authors showed that the number of vortices inside the domain $\mathcal{D}_r$ is statistically equivalent to the squared-root coarse-grained dissipation $\xi_r$ for inertial range sized contours. In this sense, the statistical repulsion among vortices at very small scales is phenomenologically related to the breakdown of multifractality as pointed out in \cite{mori1}.

Note that the $4/5$-law for circulation, as conjectured through extensive analysis of DNS data in \cite{iyer1}, which suggests $\langle|\Gamma_r|^3\rangle\sim r^4$, does not concur with the naive implementation of $\alpha=4/3$ in the VGM. This is due to a technical difference between traditional cascade models and the GMC approach. In the latter, the scaling exponents of the $p^\text{th}$-order moment do not follow the same expected relation for simple cascade models: $\langle(\sqrt{\varepsilon_r})^p\rangle\sim r^{\tau_{p/2}}$ when $\langle\varepsilon_r^p\rangle\sim r^{\tau_p}$. In the GMC approach, the evaluation of the coarse-grained dissipation rate $\varepsilon_r$ is implicitly related to the intense local fluctuation of $\varepsilon(\bm{x})$. Consequently, the only compatible way to use the GMC modeling for circulation statistics is by setting $\alpha=4/3-\mu/2$, as already pointed out in \cite{mori4}.

The procedure of cluster summation consists of characterizing the contribution of the circulation of each vortex and summing over neighboring vortices. It was first introduced in the context of quantum turbulence \cite{muller2}, where the velocity circulation is quantized, i.e., $\Gamma_i=\pm n\kappa$, with $\kappa$ being the quantum of circulation \cite{barenghi}. This procedure does not fix the circulation contour, being independent of the particularities of the local vortex distribution (or $\xi_r$ in classical turbulence), resulting in the appearance of a self-similar behavior, ultimately understood as the degree of polarization of chains of quantum vortices \cite{muller2}. 

For classical turbulence, such a summation procedure is far more involved than the quantum case. It can assume basically any value, and there may be not only sign correlations but correlations among circulation magnitudes of the vortices. Moreover, local vortex distribution is also supposed to be correlated to the vortex circulation since, at small scales, there can be decaying and/or reconnection processes occurring all over the flow. Therefore, this kind of procedure is expected to show a scaling behavior for a sufficiently large cluster. 

Suppose one has a 2D slice of a 3D turbulent flow, with $N$ vortices at positions $\bm{x}_i$ each of which carries a circulation $\Gamma_i$. We define two types of cluster summations \cite{muller2}, labeled by a function $f$, as
\be\label{clusum}
P^{(f)}_i(n)=\sum_{j=1}^{n} f\!\left(\Gamma^{(i)}_j\right)\ ,\ 
\ee
where $n\leq N$ is the size of the cluster, spanning an area with a typical size of $r\sim n^{1/2}$ due to the homogeneous behavior of the GMC scaling exponents for $p=1$. To generate the cluster summation series, we choose a specific vortex, for instance, the $i^\text{th}$ vortex. We label the circulation of each vortex as $\Gamma^{(i)}_j$ and sort them based on their relative distance $r_{ij}=|\bm{x}_i-\bm{x}_j|$ from the selected vortex. Finally, we perform a summation over the increasing $j$ label. The ensemble average is then defined by starting the series at different points $\bm{x}_i$,
\be\label{eq2p7}
\mean{P_n}=\frac{1}{N}\sum_{i=1}^N P^{(f)}_i(n)\ .\
\ee
As for the function $f:\mathbb{R}\rightarrow\mathbb{R}$, first we refer to a ``binarized'' summation when only the sign of the vortices is considered, i.e., $f(x)=\text{sign}(x)$. This basically measures the degree of polarization of the vortex cluster, exactly as in the case of quantum turbulence \cite{muller2}. Lastly, we refer to the ``continuum'' cluster summation when $f(x)=x$. Note that, however, the choice of this function instead of $|x|$ does not totally exclude the polarization effects, and, the scaling exponents of these two different series must be connected in some way. For both cases, we expect
\be\label{momsp}
\mean{\left|P_n\right|^p}\propto n^{\beta p}\ .\
\ee
If $\beta=2/3$ for $f(x)=x$, K41 scaling is achieved (since $n\sim r^2$). This result of disentangled scaling was observed in quantum turbulence to have a K41-like property as shown by \cite{muller2}, but, as mentioned, the classical case can be more involved, and a clear exploration of this scaling follows. 

Let us assume the field $\tg(x)$ is the only responsible for the scaling of the cluster summation in the VGM. Then, by the use of Eqs.~\refp{lambdap} and~\refp{momsp}, one expects for the second order moment«
\be\label{2momclus}
4\beta=4-\alpha\rightarrow \beta=\frac{2}{3}+\frac{\mu}{8}\approx 0.688\ ,\ 
\ee
so that the correction introduced in the Gaussian field $\tg(x)$ to reproduce the $4/5$-law directly influences cluster summation, i.e., the partial polarization of the vortices \cite{muller2}. At this point, we note an explicit phenomenological splitting between classical and quantum turbulence. Although the correction to $\alpha$ makes the scaling exponents of Eq.~\refp{lambdap} numerically indistinguishable from those of circulation in the quantum case \cite{iyer1,muller1}, it has completely different phenomenological bases. In the case of quantum systems, the circulation can be thought of as the product of partially polarized vortices (following Eq.~\refp{eq2p7} with $\beta=2/3$) by the local distribution of vortices which is, to first order, related to $\varepsilon_r$ as noted by \cite{muller2}. In the case of classical turbulence, \cite{mori4} noticed that the vortex density is more accurately described by $(\sqrt{\varepsilon})_r$, then to $\xi_r$. In this sense, an efficient analysis of turbulent data should elucidate the structural differences between these two systems.

\section{Data analysis}\label{sec3}

The open-access database used in this work is maintained by the Johns Hopkins University (JHTDB) \cite{JHTD,JHTD2}. We extracted 2D slices from a 3D, fully de-aliased Navier-Stokes DNS of fully developed homogeneous and isotropic turbulence, performed in a tri-periodic box of linear size $2\pi$ divided into $N=4096$ collocation points. The Taylor-based Reynolds number is about $R_\lambda\approx610$ and the integral and Kolmogorov length scales are, respectively, $L=1.3916$ and $\eta_K=1.3844\times 10^{-3}$. 

The vortex identification method we choose is the standard swirling strength criterion (SSC) which selects connected spots where $|\text{Im}(\lambda)|\geq0.125\sigma_\lambda$, with $\lambda(\bm{x})$ being any of the complex eigenvalues of the 2D velocity gradient tensor at position $\bm{x}$ ($\lambda=0$ if they are all real), and $\sigma^2_\lambda$ its variance \cite{zhou,chakraborty}. 
 
The circulation of each vortex structure can be estimated as $\Gamma_i=\omega_i^{max} A_i$, where $\omega^{max}$ is the maximum vorticity inside the detected spot and $A_i$ its area. This procedure usually overestimates circulation by discarding the effects of particular spatial decay of vorticity inside the detected area. Nevertheless, it has been shown to be a good approximation for the evaluation of circulation at large contours, provided it is not too intense (see the supplemental material of \cite{mori3}).

\subsection{Cluster Summation}

\begin{figure}[t]
\centering\includegraphics[width=\linewidth]{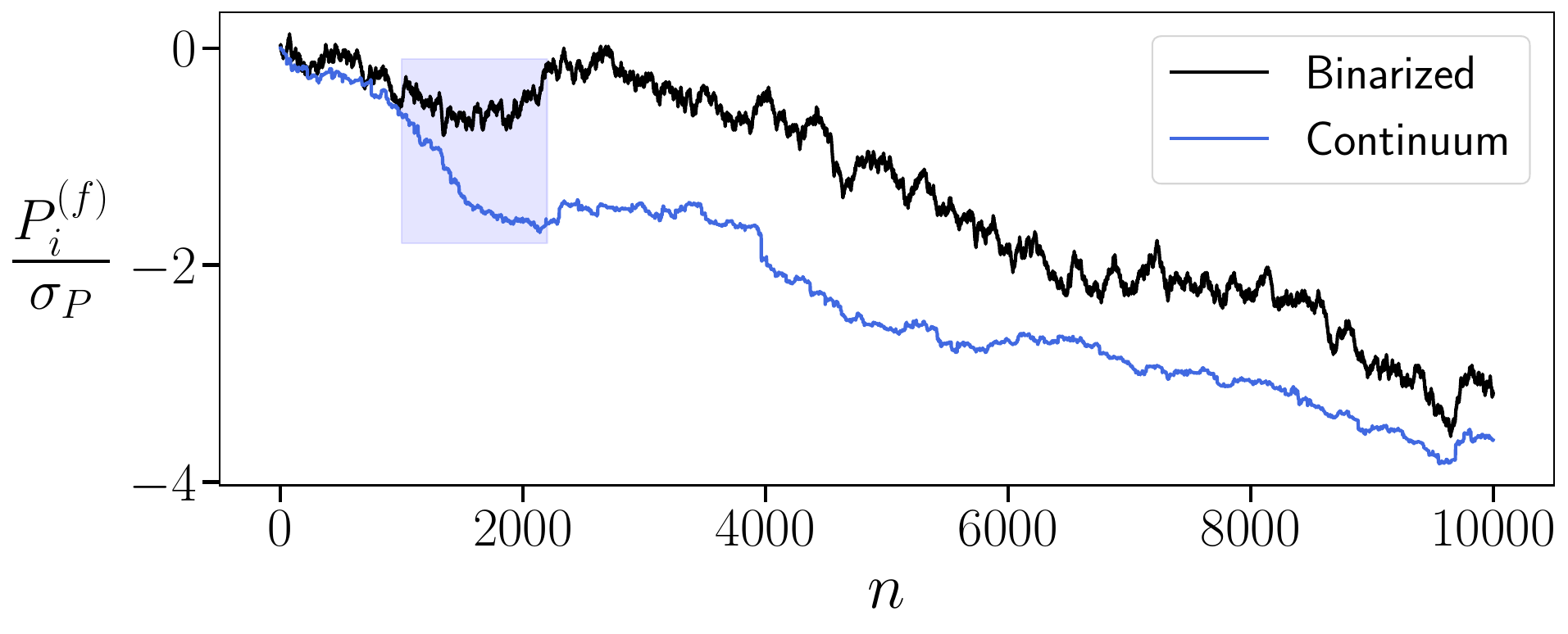}
\caption{Typical cluster summation as functions of the cluster size for both binarized (darker) and continuum (lighter) cases, for the same $\Gamma^{(i)}_j$ realization. Each realization is normalized by its own standard deviation to ease comparison. The shaded area highlights different properties of the summation methods.}\label{fig1}
\end{figure}

Fig.~\ref{fig1} shows the typical realizations of the processes defined by Eq.~\refp{clusum} as functions of the cluster size $n$. Despite showing similar large-scale behavior, small-scale peculiarities are noticeably different between the binarized and continuum cases. The effects of intensity correlations are markedly seen by taking a closer look at, for instance, the shaded region in Fig.~\ref{fig1}: one notes that the binarized polarization oscillates around a mean value, meanwhile, the continuum summation has a clear tendency to decrease. This fact indicates, in this particular case, a strong correlation among vortices with high negative intensity.

\begin{figure}[t]
\centering\includegraphics[width=\linewidth]{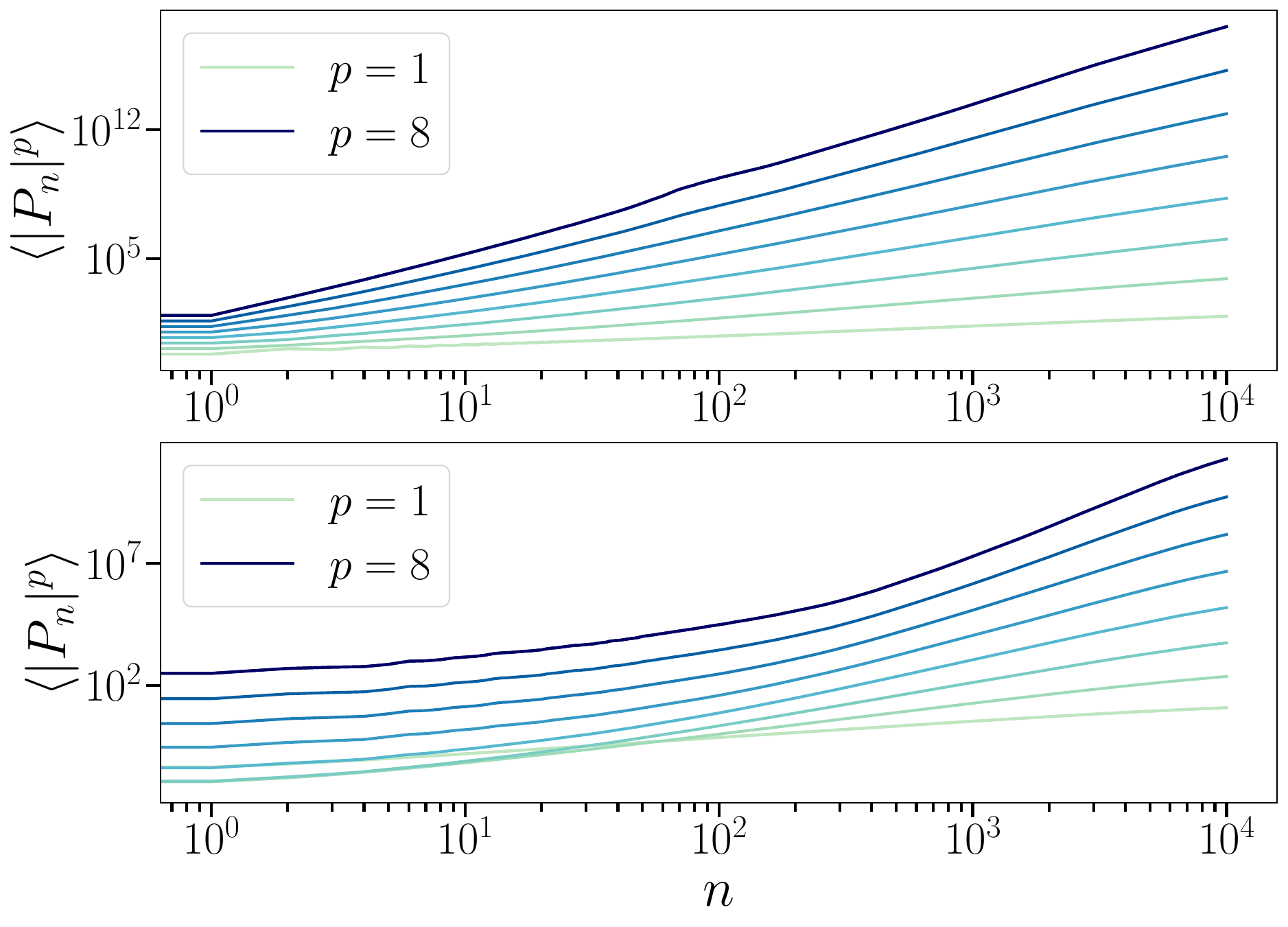}
\caption{Statistical moments of cluster summation for the binarized process (upper panel) and the continuum process (lower panel) as functions of the cluster size $n$. Statistical moments are shown from $p=1$ (lighter colors) to $p=8$ (darker colors).}\label{fig2}
\end{figure}

In Fig.~\ref{fig2}, we show the statistical moments of cluster summation in both the binarized and continuum cases, in the spirit of Eq.~\refp{momsp}.
The binarized processes show a distinct scaling in the whole cluster size range studied $(n\leq10^4)$, resulting in a general scaling exponent given by Eq.~\refp{momsp} with $\beta=0.56\pm0.01$. This means that the tangle of point vortices in classical turbulence is less polarized than in quantum turbulence ($\beta=2/3$). In addition, not only the polarization of the vortices, but their amplitude correlations are important for the K41 scaling, so that the full scaling of the continuum process can be more fundamental than the sole polarization of the vortices.

In striking contrast to its binarized version, the continuum cluster summation presents no clear scaling in the whole cluster size range. However, for very large clusters one can identify a scaling regime depending on the moment order, such that the higher the moment order, the shorter the scaling range.
The determination of the power law exponent in the large cluster scaling range is difficult because of the substantial contamination due to the slow crossover between small and large-scale behaviors. Moreover, a range of about $10^4$ vortices has, in general, a spanning area comparable to integral length scales, so the statistics can start to be affected by the energy pumping mechanism, again, affecting the scaling region by further shortening it, now from the right side of Fig.~\ref{fig2}.

\subsection{Small Cluster Regularization}

One can try to understand the scaling range by noting that the behavior of the cluster summation at very small contours is affected by the regularization of $\tg(\bm{x}\rightarrow 0)$. Indeed, this was the leading mechanism used in \cite{apol1} for the saturation of the kurtosis for $r\sim \eta_K$. 

To further illustrate this point, we assume that, within the VGM framework, the cluster summation is solely determined by the contribution of $\tg(\bm{x})$, viz.,
\be\label{eq3p1}
S_n=\sum_{i=1}^{n}\tg(\bm{x}_i) \ . \
\ee
Writing the variance of $S_n$, one can split the contributions into
\be
\mean{S_n^2}=\sum_{i=1}^n\mean{\tg^2 (\bm{x}_i)}+2\sum_{i=1}^n\sum_{j=i+1}^n\mean{\tg(\bm{x}_i)\tg(\bm{x}_j)} \ .\
\ee
Assuming now that the $\tg$ field is regularized, i.e., $\langle\tg( \bm{x}_i)\tg( \bm{x}_j)\rangle=\Gamma_0^2$ for $i=j$ and $\langle\tg( \bm{x}_i)\tg( \bm{x}_j)\rangle=\Gamma_0^2| \bm{x}_i-  \bm{x}_j|^{-\alpha}$ for $i\neq j$, the cluster summation's variance reads
\be\label{combined}
\mean{S_n^2}=\Gamma_0^2\left(n+\frac{4}{2-\alpha}(\pi\bar\sigma)^{\alpha/2} n^{(4-\alpha)/2}\right) \ ,\
\ee
where $\bar\sigma=N/L^2$ is the mean vortex density, whose value, obtained numerically, is $\bar\sigma\approx 3\times 10^{-3}$ in lattice units. By the usage of Eq.~\refp{2momclus}, the asymptotically dominant scaling exponent is $(4-\alpha)/2=2\beta$ as expected. For our purposes, we need a sufficiently large cluster, say, $n\gg n_0$ with
\be\label{n0}
n_0=\left(\frac{4}{2-\alpha}(\pi\bar\sigma)^{\alpha/2}\right)^{2/(\alpha-2)}\approx 27 \ ,\
\ee
for the self-similar exponent to dominate, since $(4-\alpha)/2>1$. Unfortunately, the structures are expected to be less dense for higher Reynolds numbers, which brings several difficulties for the scaling of cluster summation. One may expect, for instance, that the cross-over point $n_p$ between small-scale and large-scale behavior of the cluster summation statistical moments should grow with the moment order $p$. This can be grasped by assuming a Gaussian behavior for $\tg(x)$, from where one may apply Isserlis' theorem \cite{isserl} to arrive at
\begin{figure}[b]
\centering\includegraphics[width=\linewidth]{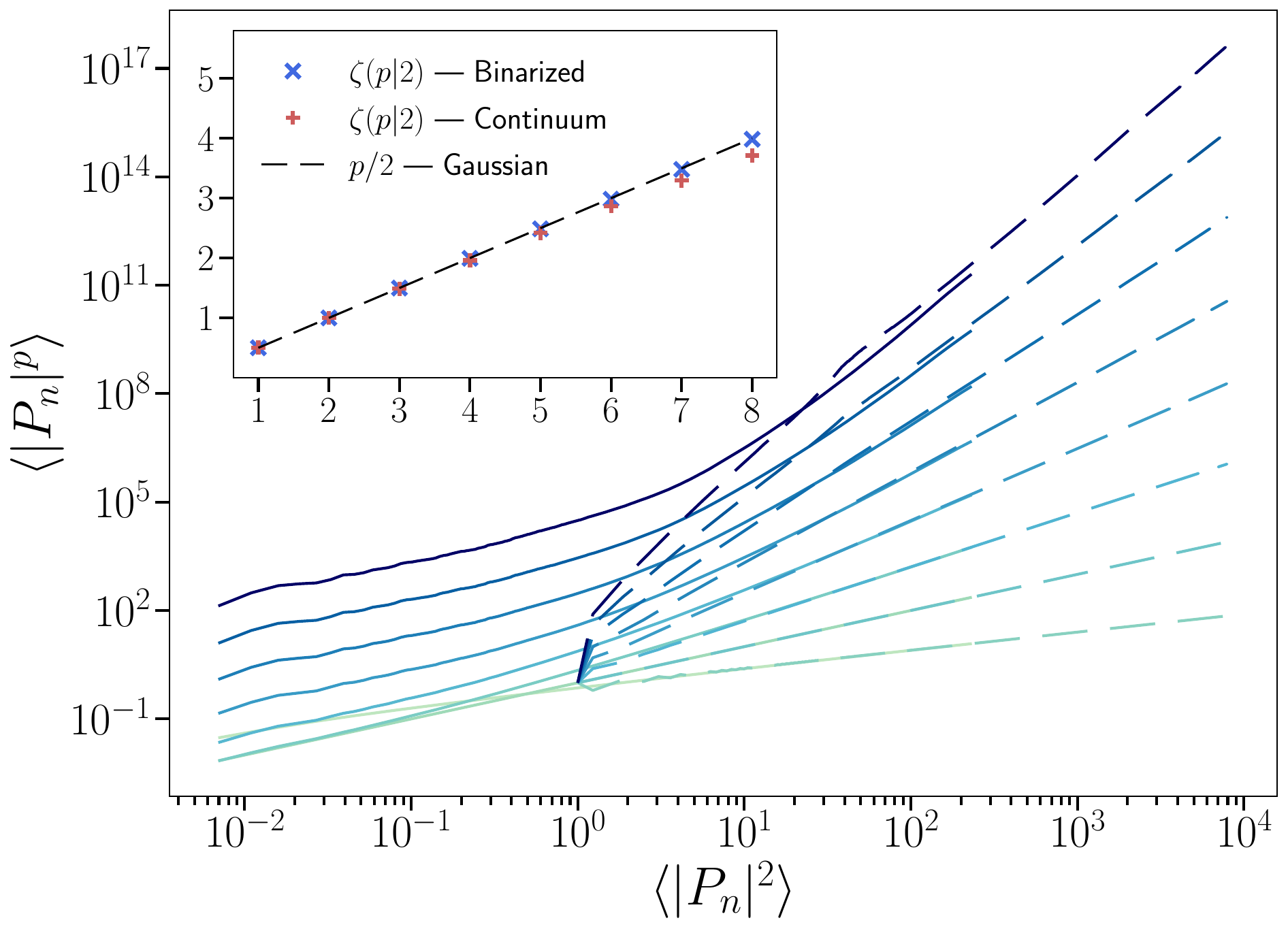}
\caption{ESS approach for the cluster summation scaling on the binarized (dashed) and continuum (solid) cases. Colormaps are as in Fig.~\ref{fig2}. Inset: ESS exponents found on both cases compared to the Gaussian values $p/2$ (dashed).}\label{fig3}
\end{figure}
\be\label{smomsp}
\mean{S_n^{2p}}\sim\left(n+\frac{4}{2-\alpha}(\pi\bar\sigma)^{\alpha/2} n^{(4-\alpha)/2}\right)^p \ ,\
\ee
which reduces to
\be
\mean{S_n^{2p}}\propto n^{2\beta p}\left[1+\lr{\frac{n_p}{n}}^{(2\beta-1)}+\mathcal{O}\lr{\lr{\frac{1}{n^{2\beta-1}}}^2}\right]\ ,\
\ee
with $n_p\equiv n_0 p^{1/(2\beta-1)}$. Considering Eq.~\refp{2momclus}, one has $n_p=n_0 p^{\xi}$, with $\xi=12/(4+3\mu)\approx2.66$. In short words, this crossover would grow as a power law, due to the presence of the small-scale regularization. However, the presence of highly correlated small-scale clusters plays an important role in the crossover and the Gaussian behavior over the whole range of $n$'s is known to be false. Indeed, the prediction for the eighth-order moment $n_8\approx6800$ is unrealistic, as seen in Fig.~\ref{fig2}, and hence this analysis is simply indicative.

\subsection{Scaling Exponents and Asymptotic Gaussianity}

An extended self-similarity (ESS) \cite{ESS} analysis was carried out for both the binarized and continuum scalings and results are shown in Fig.~\ref{fig3}. In the ESS approach, one expects that although the scaling of the statistical moments as a function of $n$ may not be clear-cut, the $p^\text{th}$ order moment has the same small-scale behavior as the $q^\text{th}$ order moment, such that, for the particular case $q=2$, $\mean{|P_n|^q}\propto\mean{|P_n|^2}^{\zeta(p|2)}$. This is a usual strategy of the ESS approach to extend the scaling region and get better-resolved exponents. Nonetheless, as shown in Fig.~\ref{fig3}, the continuum process does not exhibit this property due to the collapsing moments for $n\rightarrow0$ as emphasized by Eq.~\refp{smomsp}.

A signature of Gaussian behavior in the binarized summation can be found on the inset of Fig.~\ref{fig3}, where the measured ESS exponents are depicted. Despite a slight curvature for high-order moments, the Gaussian behavior for the continuum cluster summation is restored at larger cluster sizes. One may note, as suggested by Fig.~\ref{fig3}, that the scaling of the continuum process tends asymptotically to the binarized one, such that self-similarity is restored for sufficiently big clusters.
\begin{figure}[b]
\centering\includegraphics[width=\linewidth]{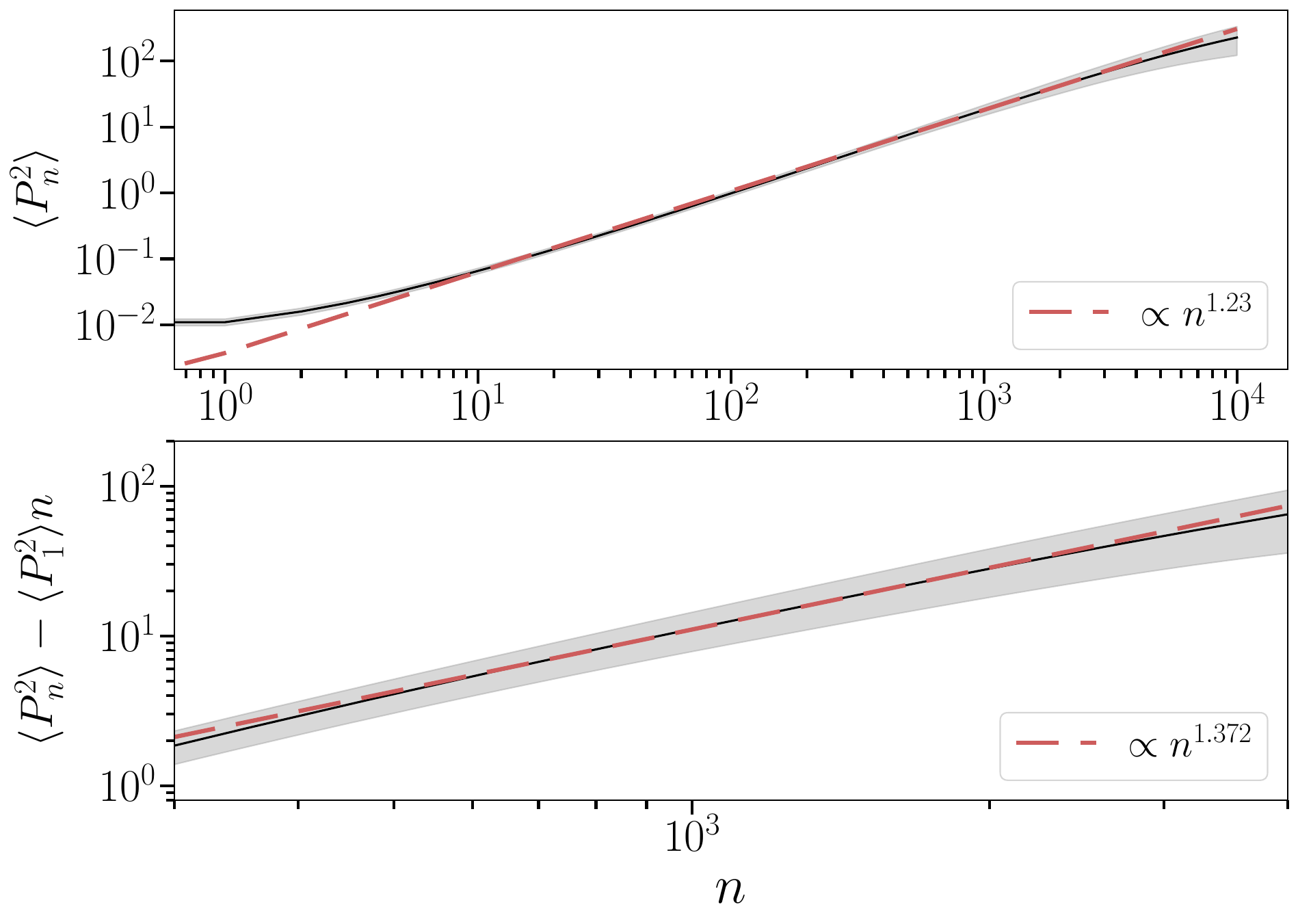}
\caption{Continuum cluster summation's variance is shown in the upper frame, while its subtracted version is shown in the bottom frame. Dashed lines are power-law fits while shaded region represents a $2\sigma$-confidence interval of the data.}\label{fig4}
\end{figure}
\begin{figure}[t]
\centering\includegraphics[width=\linewidth]{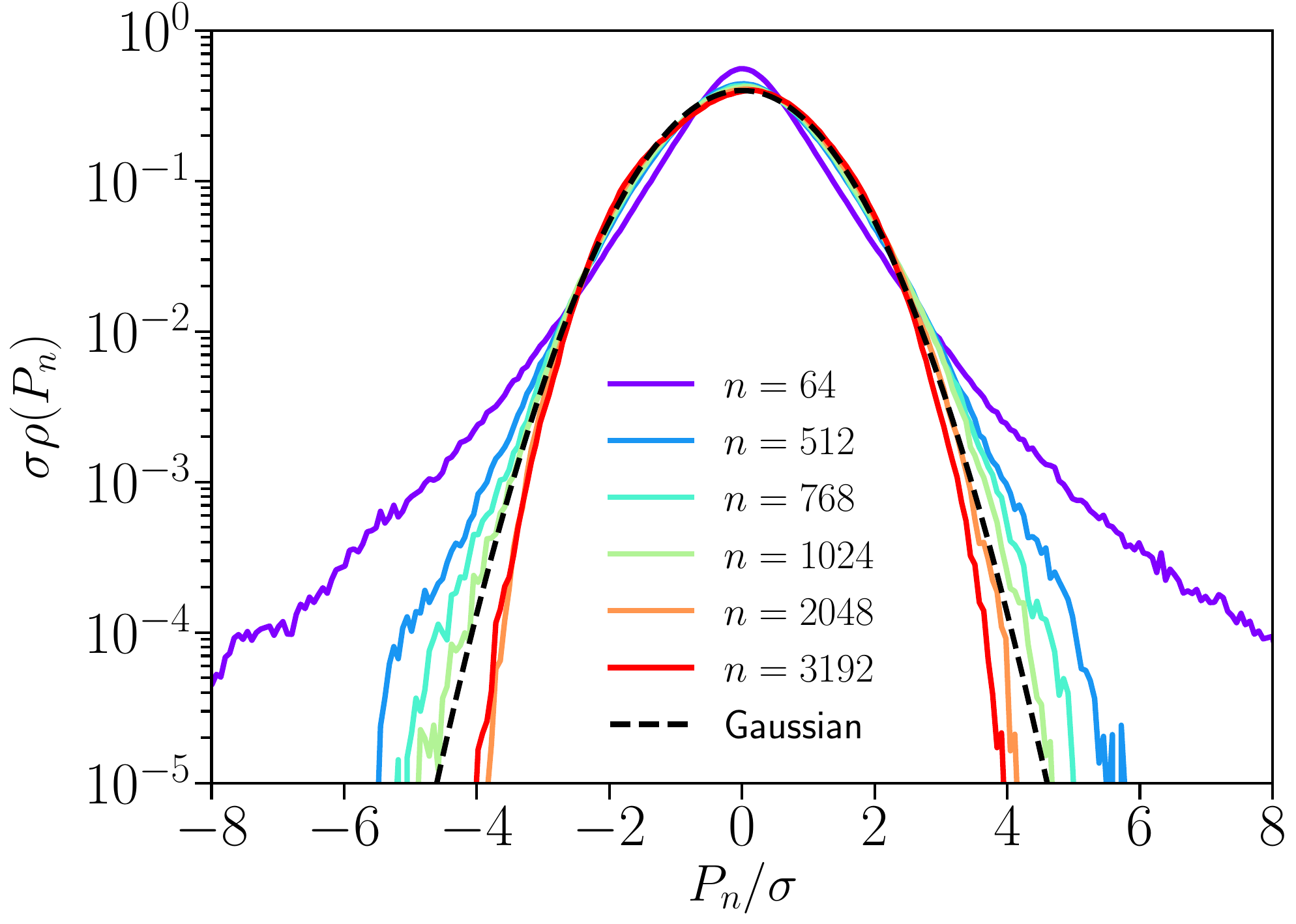}
\caption{Standardized circulation PDFs of cluster summation for various cluster sizes.}\label{fig5}
\end{figure}

As a next step in fully characterizing the self-similar behavior of the cluster summation, one needs to determine the scaling exponent of one of its statistical moments, which we take, for simplicity, $p=2$, since $\zeta(p|2)$ was already computed. Furthermore, if one interprets the cluster summation as Eq.~\refp{eq3p1} in the VGM, caution must be taken due to an unexpected numerical coincidence between the power law $n^{5/4}$ and the curve provided by Eq.~\refp{combined} for the parameters $\bar\sigma$, $\alpha$ and $\Gamma_0$ in our data. This exponent roughly fits the behavior of the variance as it may be inferred from Fig.~\ref{fig4} (upper panel), where a similar scaling is depicted ($\sim n^{1.23}$).

The lower panel of Fig.~\ref{fig4} shows a properly subtracted scaling, following Eq.~\refp{combined}. The scaling of the observables $\left(\mean{P^2_n}-\mean{P^2_1}n\right)$ and $\mean{P^2_n}$ are compatible with power laws proportional to $n^{2\beta}$, with $\beta=0.686\pm0.018$ and $\beta=0.615\pm0.011$, respectively. 

Further evidence supporting the Gaussian behavior of the large-scale continuum summation is found in Fig.~\ref{fig5}. Clusters achieve nearly Gaussian PDFs in the range $n\approx$ 700--2000 where the exponents were fitted.

\subsection{Spatial Correlation Function}

\begin{figure}[t]
\centering\includegraphics[width=\linewidth]{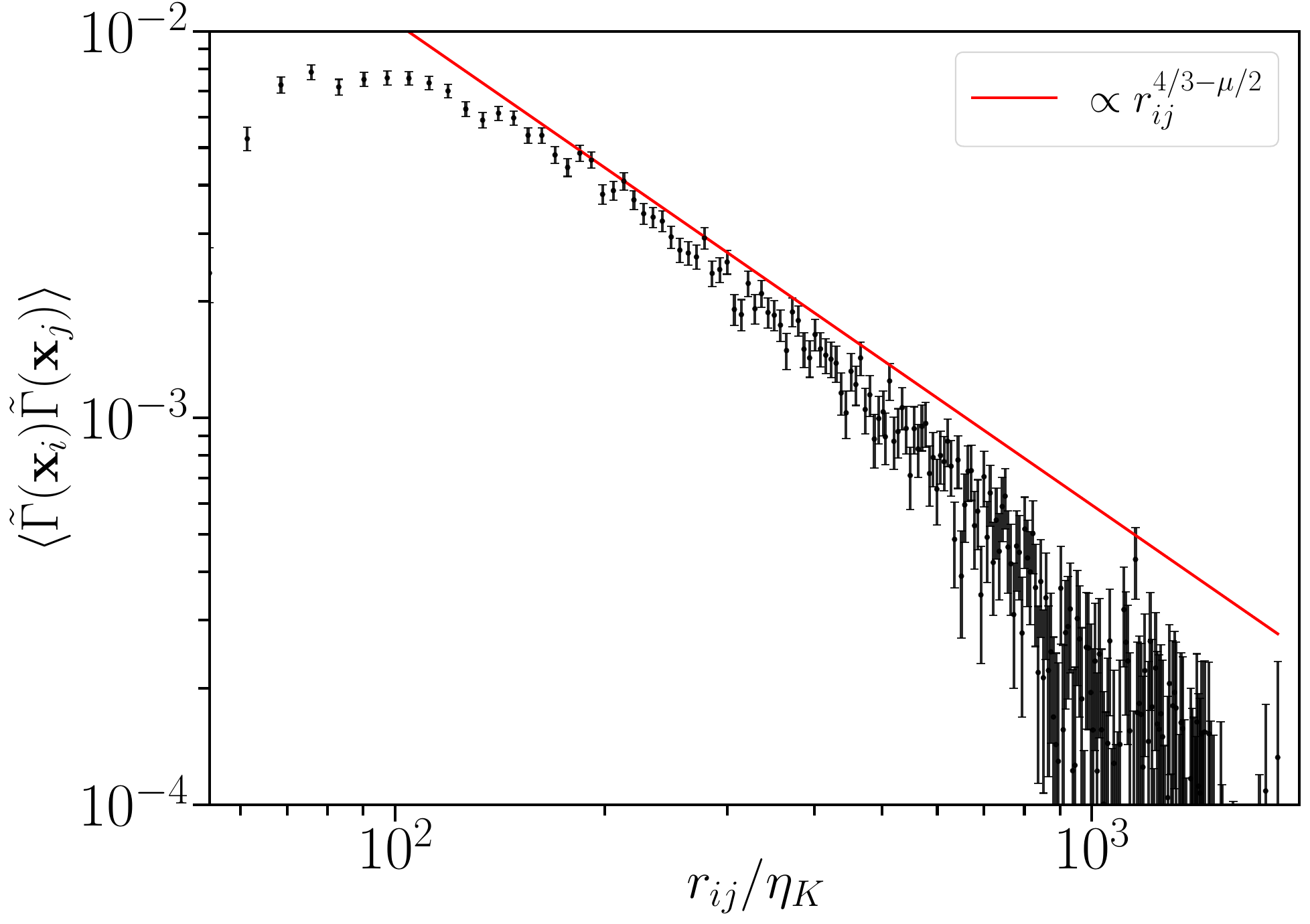}
\caption{Two-point correlation of the detected vortex circulation as a function of the inter-vortex distance.}
\label{fig6}
\end{figure}

To further validate the cluster summation methodology, we show in Fig.~\ref{fig6} the vortex circulation two-point correlation function, $\langle \tg(\bm{x}_i)\tg(\bm{x}_j)\rangle$, as a function of the inter-vortex distance $r_{ij}=|\bm{x}_i-\bm{x}_j|$, directly measured from DNS data. 
To achieve this, we measured individual vortices circulations, then took pairwise products $\tilde\Gamma(\bm{x}_i)\tilde\Gamma(\bm{x}_j)$ on each snapshot and computed averages over bins organized according to $r_{ij}$. Results are considerably robust with regard to bin size, so we choose it as twice the lattice parameter $dx$ of the DNS data, corresponding to $dx\approx 1.1\eta_K$. Error bars on the figure correspond to one standard error of the mean.

While the high amount of fluctuations does not allow one to distinguish a precise scaling, the fitted exponent $\alpha=1.31\pm0.10$ turns out to be compatible with the VGM prediction (shown as a solid line).

\section{Discussion}\label{sec4}

Previous explorations of the building blocks of the VGM have revealed important structural properties of the general statistical behavior of circulation statistics in classical fully developed turbulence \cite{mori2,mori4}. The so-far unexplored phenomenological field introduced in the VGM, $\tg(\bf x)$, was considered in this work. The current findings start to signal some fundamental differences in the building blocks of circulation statistics in classical and quantum turbulence. By exploring the degree of vortex polarization through the binarized cluster summation method, we found clearly incompatible scaling exponents for these two systems, where $\mean{P^2_n}\sim n^{1.120\pm0.022}$ stands for the classical case, while $\mean{P^2_n}\sim n^{4/3}$ to the quantum one. Moreover, we found power law exponents to be compatible with the correction predicted by the GMC approach, introduced in the VGM as $\langle \tg(\bm{x})\tg(\bm{y})\rangle\sim n^{2\beta}$ with $\beta=2/3+\mu/8\approx0.688$, to be in complete agreement with the measured value $\beta=0.686\pm0.018$ through the evaluation of the continuum cluster summation.

The main structural similarities and differences between the modeling constituents of quantum and classical turbulence can be summarized as: (i) both systems can be modeled through the product of the partial polarization (measured by the cluster summation) and the spatial distribution of vortices (which brings the intermittency of circulation through the energy dissipation ratio); (ii) vortex distribution in the classical system is related to the field $\sigma\sim\xi_r\sim(\sqrt{\varepsilon})_r$, while, to first approximation, in the quantum system it is related to $\sigma_r\sim\varepsilon_r$; (iii) the partial polarization of the quantum system displays K41 behavior, which is broken in the classical system due to strong fluctuations in individual vortex intensities, while remaining self-similar for big cluster sizes; (iv) circulation in both systems displays the same overall scaling $\lambda_p^{classical}=\lambda_p^{quantum}$ at inertial range scales. At this point, the rationale brought by the combination of points (i), (iii), and (iv) is that the spatial distribution of vortices must be different when comparing both systems. However, the connection of this conclusion to point (ii) evokes a deeper understanding of the specific roles of the underlying dynamics of Navier-Stokes and Gross-Pitaevskii equations on the spatial distribution of vortices and how this is connected to the local energy dissipation rate.

For the models based on the standard machinery of random cascades, such as the mOK62 introduced by \cite{muller2}, the correction to the cluster summation is not present since the scaling exponents of the coarse-grained dissipation exactly vanish for $p=3$. On the other hand, the best bifractal fit achieved by \cite{iyer1} for low order moments $\lambda_{p}\approx (1.367\pm0.009)p$ was referred to as $1.4$ scaling, since the exponent $1.4$ collapses the circulation PDFs cores. In the present work, we associate this value not only to the VGM's prediction but to the correction introduced by the GMC field, $\lambda_{p\rightarrow 0}\approx(4/3+3\mu/8)p=1.397p$.

Recently, experimental measures of the circulation scaling exponents in quasi-two-dimensional turbulence were reported in \cite{zhu}. Under the inverse energy cascade regime, circulation shows very similar --not to say equal-- scaling behavior as standard three-dimensional homogeneous and isotropic turbulence. These different systems, having similar overall statistical properties when analyzed through the magnifying glass of the circulation variable, call attention to the possibility of a broader unifying theory for circulation statistics. The latter motivates the search for an extension of the VGM at different phenomenological systems such as rotating turbulence or magnetohydrodynamics, where the existence of a preferential direction breaks isotropy. In this sense, the VGM setting and the role of a bounded GMC in the vortex distribution can shed some light on the problems of clustering, coalescence, and polarization of structures at different turbulent systems.

\begin{acknowledgments}
V.J.V. thanks to G. Krstulovic and N. M\"uller for enlightening discussions about the cluster summation method. This work was partially supported by CAPES (grant n$^o$ 88887.336246/2019-00) and CNPq.
\end{acknowledgments}

\end{document}